\documentclass[12pt]{article}
\usepackage{amsmath}
\def\be{\begin{equation}}
\def\ee{\end{equation}}
\def\arr{\begin{array}{rll}}
\def\ea{\end{array}}
\def\bea{\begin{eqnarray}}
\def\eea{\end{eqnarray}}

\def\N2{$N{=}2$}

\def\>{\rangle}
\def\<{\langle}
\def\+{\dagger}
\def\={\ =\ }

\begin{document}
\renewcommand{\thefootnote}{\fnsymbol{footnote}}
\begin{titlepage}
\setcounter{page}{0}

\vskip 1cm
\begin{center}
{\LARGE\bf  Remark on integrable deformations   }\\
\vskip 0.5cm
{\LARGE\bf of the Euler top }\\
\vskip 2cm
$
\textrm{\Large Anton Galajinsky \ }^{a,b}
$
\vskip 0.7cm
${}^{a}$
{\it
Laboratory of Mathematical Physics, Tomsk Polytechnic University, \\
634050 Tomsk, Lenin Ave. 30, Russian Federation} \\
\vskip 0.5cm
${}^{b}$
{\it
Department of Physics, Tomsk State University, 634050 Tomsk, \\
Lenin Ave. 36, Russian Federation} \\
{Email: galajin@tpu.ru}

\end{center}
\vskip 1cm
\begin{abstract}
\noindent
The Euler top describes a free rotation of a rigid body about its center of mass and provides an important example of a completely integrable system. A salient feature of its first integrals is that, up to a reparametrization of time, they uniquely determine the dynamical equations themselves. In this note, this property is used to construct integrable deformations of the Euler top.
\end{abstract}

\vspace{0.5cm}

\end{titlepage}

\renewcommand{\thefootnote}{\arabic{footnote}}
\setcounter{footnote}0

The Euler top describes a free rotation of a rigid body about its center of mass.
Apart from its role in analytical mechanics, it provides an important example of
a completely integrable system.
The equations of motion\footnote{These equations follow from the original Euler equations by a scaling transformation $x_1 \rightarrow a_1 x_1$, $x_2 \rightarrow a_2 x_2$, $x_3 \rightarrow a_3 x_3$, with properly chosen number coefficients $a_1,a_2,a_3$.
}
\be\label{Ee}
\dot x_1=x_2 x_3, \qquad \dot x_2=x_1 x_3, \qquad \dot x_3=x_1 x_2,
\ee
where $x_i$ are interpreted as components of the angular velocity vector,
admit two functionally independent first integrals
\be\label{fi}
x_1^2-x_2^2=C_1, \qquad x_1^2-x_3^2=C_2.
\ee
These can be used to express any two components in terms of the third one. Substitution in (\ref{Ee}) then yields a single ordinary differential equation, which is solved by quadrature. The general solution is given in terms of elliptic functions.
Note that the Euler equations can be put into the Hamiltonian form, in which they describe a motion on the group space of $SO(3)$ \cite{Ar}. This observation is a clue to the construction of higher dimensional integrable generalizations \cite{Man} (see also \cite{Far}--\cite{F2}).

A salient feature of the first integrals (\ref{fi}) is that, up to a reparametrization of time, they uniquely determine the dynamical equations themselves. Indeed, the derivative of (\ref{fi}) yields
\be
\dot x_1=\frac{x_3}{x_1} \dot x_3, \qquad \dot x_2=\frac{x_3}{x_2} \dot x_3,
\ee
with $\dot x_3$ being unspecified. Setting
\be
\dot x_3=x_1 x_2 f,
\ee
where $f=f(x_1,x_2,x_3)$ is an arbitrary function, and redefining the temporal coordinate
\be
t \quad \rightarrow \quad s=\int_0^t f(\tau) d\tau,
\ee
one arrives at (\ref{Ee}), where the dot now designates the derivative with respect to the new parameter $s$.

The purpose of this brief note is to use this feature to construct integrable deformations of the Euler top.

Let us alter the first integrals (\ref{fi}) by adding to them two arbitrary differentiable functions $\alpha=\alpha (x_1,x_2,x_3)$
and $\beta=\beta (x_1,x_2,x_3)$
\be\label{fi1}
x_1^2-x_2^2+2 \alpha=C_1, \qquad x_1^2-x_3^2+2 \beta=C_2.
\ee
Repeating the steps above, one finds the system of ordinary differential equations ($\partial_i=\frac{\partial}{\partial x_i}$)
\bea\label{de}
&&
\dot x_1=x_2 x_3+\partial_2 \alpha \partial_3 \beta-\partial_3 \alpha \partial_2 \beta-x_3 \partial_2 \alpha-x_2 \partial_3 \beta,
\nonumber\\[2pt]
&&
\dot x_2=x_1 x_3+\partial_3 \alpha \partial_1 \beta-\partial_1 \alpha \partial_3 \beta+x_1 \partial_3 (\alpha-\beta)+x_3 \partial_1 \alpha,
\nonumber\\[2pt]
&&
\dot x_3=x_1 x_2+\partial_1 \alpha \partial_2 \beta-\partial_2 \alpha \partial_1 \beta-x_1 \partial_2 (\alpha-\beta)+x_2 \partial_1 \beta,
\eea
for which (\ref{fi1}) are the first integrals.

A few comments are in order. First, the equations (\ref{Ee}) describe a free rotation of a rigid body. It seems natural to interpret the extra terms in (\ref{de}) as due to a torque. Second, although the system (\ref{de}) is Liouville integrable, it is not always exactly solvable. In general,
(\ref{fi1}) can not be resolved to express two variables from the triple $(x_1,x_2,x_3)$ as explicit functions of the third one. Third, in some instances (\ref{de}) can be linked to (\ref{Ee}) by a coordinate transformation. If the equations\footnote{That (\ref{fi1}) are the first integrals of (\ref{de}) imply that the new variables obey $\dot X_1=\frac{X_2}{X_1} \dot X_2$, $\dot X_3=\frac{X_2}{X_3} \dot X_2$.}
\be\label{sup}
x_1^2-x_2^2+2 \alpha=X_1^2-X_2^2, \qquad  x_1^2-x_3^2+2 \beta=X_1^2-X_3^2,
\ee
where $X_1,X_2,X_3$ are the new variables, can be solved, for instance, for $x_1$ and $x_3$, then the transformation law for $x_2$
can be found by substituting $x_2=F(X_1,X_2,X_3)$ into the second line in (\ref{de}) and demanding $\dot X_2=X_1 X_3$. This gives
a first order linear inhomogeneous partial differential equation for $F(X_1,X_2,X_3)$. Note that, in general, $F(X_1,X_2,X_3)$ can be found as an implicit function only.  Fourth, as written in (\ref{Ee}) and (\ref{fi}), the Euler equations and the first integrals are invariant under the permutations of any pair from the triple $(x_1,x_2,x_3)$.
In general, the deformation (\ref{de}) does not maintain this property. It is instructive to give a couple of examples in which both the equations of motion and the first integrals respect the permutation symmetry.
The simplest choice
\be\label{fi2}
x_1^2-x_2^2+\frac{2g}{x_1}-\frac{2g}{x_2}=C_1, \qquad x_1^2-x_3^2+\frac{2g}{x_1}-\frac{2g}{x_3}=C_2,
\ee
where $g$ is treated as a deformation parameter (a coupling constant), yields
\bea
&&
\dot x_1=x_2 x_3 -\frac{g (x_2^3+x_3^3)}{{(x_2 x_3)}^2}+\frac{g^2}{{(x_2 x_3)}^2}, \qquad \dot x_2=x_1 x_3 -\frac{g (x_1^3+x_3^3)}{{(x_1 x_3)}^2}+\frac{g^2}{{(x_1 x_3)}^2},
\nonumber\\[2pt]
&&
\dot x_3=x_1 x_2 -\frac{g (x_1^3+x_2^3)}{{(x_1 x_2)}^2}+\frac{g^2}{{(x_1 x_2)}^2}.
\eea
Note that, as far as explicit integration is concerned, (\ref{fi2}) imply cubic algebraic equations. A further example gives a system in which first integrals amount to a coupled set of quartic algebraic equations
\bea
x_1^2-x_2^2+\frac{g}{x_{12} x_{13}}-\frac{g}{x_{21} x_{23}}=C_1, \qquad x_1^2-x_3^2+\frac{g}{x_{12} x_{13}}-\frac{g}{x_{31} x_{32}}=C_2,
\eea
where $x_{ij}=x_i-x_j$. These yield
\bea
&&
\dot x_1=x_2 x_3+\frac{g[x_1(x_2+x_3)-2 x_2 x_3]}{2 x_{12} x_{13} x_{23}^2}+\frac{g [x_2^2+x_3^2-x_1(x_2+x_3)]}{x_{12}^2 x_{13}^2}-\frac{3 g^2}{2 x_{12}^2 x_{13}^2 x_{23}^2},
\nonumber\\[2pt]
&&
\dot x_2=x_1 x_3+\frac{g[x_2(x_1+x_3)-2 x_1 x_3]}{2 x_{21} x_{23} x_{13}^2}+\frac{g [x_1^2+x_3^2-x_2(x_1+x_3)]}{x_{12}^2 x_{23}^2}-\frac{3 g^2}{2 x_{12}^2 x_{13}^2 x_{23}^2},
\nonumber\\[2pt]
&&
\dot x_3=x_1 x_2+\frac{g[x_3(x_1+x_2)-2 x_1 x_2]}{2 x_{32} x_{31} x_{12}^2}+\frac{g [x_1^2+x_2^2-x_3(x_1+x_2)]}{x_{23}^2 x_{13}^2}-\frac{3 g^2}{2 x_{12}^2 x_{13}^2 x_{23}^2}.
\eea
Both the examples above reduce to the Euler top in the limit in which the deformation parameter $g$ tends to zero. Fifth, a natural generalization of the Euler top to the case of $n$--dimensional space reads \cite{Far}
\be
\dot x_i=\prod_{j \ne i}^n x_j, \qquad x_1^2-x_2^2=C_1, \quad \dots \quad, \quad x_1^2-x_{n}^2=C_{n-1},
\ee
where $i,j=1,\dots,n$.
The analysis above is immediately applicable to this case as well.

\vspace{0.5cm}

\noindent{\bf Acknowledgements}\\

\noindent
This work was supported by the RFBR grant 13-02-90602-Arm.

\end{document}